\documentclass[12pt]{article}
\usepackage[utf8]{inputenc}

\usepackage{comment}
\usepackage{amsmath}
\usepackage{float}
\usepackage{subcaption}
\usepackage{setspace}
\usepackage[ruled,linesnumbered]{algorithm2e}
\usepackage{xcolor}
\usepackage{graphicx,amsfonts,amsthm,amssymb}
\usepackage[colorlinks,citecolor=blue,urlcolor=blue]{hyperref}
\usepackage{natbib}
\usepackage{amsmath,mathrsfs, bm}
\usepackage{wrapfig,lipsum,booktabs}
\usepackage{appendix}
\usepackage{enumitem}
\usepackage{multirow}
\usepackage{natbib}
\usepackage{cleveref}
\newtheorem{definition}{Definition}
\usepackage{enumitem}

\newcommand{\blind}{1}

\def\spacingset#1{\renewcommand{\baselinestretch}%
{#1}\small\normalsize} \spacingset{1}

\newcommand{\pr}{\mathbb{P}}

\newcommand{\Ex}{\mathbb{E}}
\newcommand{\1}{{\rm 1}\kern-0.24em{\rm I}}
\newcommand{\R}{{\rm I}\kern-0.18em{\rm R}}

\newcommand{\cS}{\mathcal{S}}

\newcommand{\bX}{\mathbf{X}}

\newcommand{\by}{\mathbf{y}}
\newcommand{\bZ}{\mathbf{Z}}

\usepackage[bottom]{footmisc}
\newtheorem{theorem}{Theorem}
\newtheorem{example}{Example}
\newtheorem{assumption}{Assumption}

\newtheorem{simsetting}{Simulation Setting}

\def\bSigma{\boldsymbol\Sigma}

\def\bbeta{\boldsymbol\beta}

\def\bvarepsilon{\boldsymbol\varepsilon}

\usepackage[bottom]{footmisc}

\def\pr{\mathbb{P}}
\def\Ex{\mathbb{E}}

\def\bSigma{\boldsymbol\Sigma}

\def\CI{\mathrm{CI}}

\renewcommand{\appendix}{
 \setcounter{section}{0}%
  \setcounter{subsection}{0}%
  \renewcommand\thesection{\Alph{section}}
  \setcounter{equation}{0}
  \renewcommand{\theequation}{S.\arabic{equation}}
  \setcounter{figure}{0}
  \renewcommand\thefigure{S\arabic{figure}}
  \setcounter{table}{0}
  \renewcommand\thetable{S\arabic{table}}  
  }

\allowdisplaybreaks

\begin{document}

\if1\blind
{
  \title{
  TwinPoSI: Statistical Twins for Powerful Post-Selection Inference on Population-Based Regression Coefficients
  }
   \author{ 
   Changhu Wang\thanks{Equal contribution} \\
  Biostatistics
Program, Public Health Science
Division \\ Fred Hutchinson Cancer Center \\
  Zhifei Li\footnotemark[1] 
  \\
   School of Mathematical Sciences, Peking University \\
   Jingyi Jessica Li\thanks{Correspondence should be addressed to Jingyi Jessica Li (jli@stat.ucla.edu, lijy03@fredhutch.org)}\\
 Biostatistics
Program, Public Health Science
Division \\ Fred Hutchinson Cancer Center
}
  \date{}
  \maketitle
} \fi

\if0\blind
{
  \bigskip
  \bigskip
  \bigskip
  \begin{center}
    {\LARGE\bf Title}
\end{center}
  \medskip
} \fi

\maketitle

\def\spacingset#1{\renewcommand{\baselinestretch}%
{#1}\small\normalsize} \spacingset{1}


\begin{abstract}
Post-selection inference (PoSI) quantifies uncertainty about regression coefficients after data-driven feature selection. A central question is the inferential target. Projection-based targets are the coefficients of the best linear approximation restricted to the selected features and depend on the selection procedure, whereas population-based targets are the selected features' coefficients in the full linear model and are often more relevant for scientific interpretation. The two targets can differ, even in sign. Existing PoSI methods largely focus on projection-based targets and may yield overly wide confidence intervals. We propose TwinPoSI (statistical twin post-selection inference), a method for powerful PoSI on population-based targets with false coverage rate (FCR) control. TwinPoSI leverages \textit{statistical twins}, defined as synthetic datasets generated from a fitted model to mimic the original data. With generating parameters known, statistical twins enable direct calibration of confidence intervals for bias introduced by data-dependent feature selection. We establish that TwinPoSI controls the FCR at the nominal level, up to Monte Carlo error and an asymptotically negligible approximation error. Moreover, in a Gaussian linear model with an orthogonal design, TwinPoSI uniquely attains the oracle average interval length to first order among the methods considered. Simulations show that TwinPoSI maintains reliable FCR control for both population-based and projection-based targets while yielding shorter confidence intervals and higher power than the competing methods. In riboflavin production data, TwinPoSI identifies associated genes, quantifies their directions and magnitudes of association, and yields the shortest average intervals among the methods considered.
\end{abstract}

\newpage

\spacingset{1.9}

\section{Introduction}
In the current data analysis workflow, data preprocessing is often a prerequisite for statistical inference \citep{Kuchibhotla2021PostselectionI}; for instance, in high-dimensional data analysis, researchers typically prioritize selecting important features to circumvent the dimensionality crisis, and then perform hypothesis testing or construct confidence intervals for the selected features. Classical statistical inference methods, however, do not account for the bias introduced by the previous data-dependent feature selection step, which can lead to invalid inference. This issue is commonly referred to as selection bias in post-selection
inference (PoSI) \citep{berk2013valid}, which can inflate the false discovery rate (FDR), defined as the expected
proportion of false positives among reported discoveries \citep{benjamini1995controlling}, and the false coverage
rate (FCR), the expected proportion of noncovering intervals among the
reported confidence intervals
\citep{benjamini2005false}. Such inflation can lead to unreliable or
misleading scientific conclusions. In this paper, we focus on the PoSI problem of constructing confidence intervals (CIs) for the coefficients associated with the selected features.

We consider the following high-dimensional linear regression model:
$
\by = \bX \bbeta^* + \bvarepsilon,
$
where $\by \in \mathbb{R}^n$ denotes the response vector, and $\bX \in \mathbb{R}^{n \times p}$ is the fixed design matrix, whose rows correspond to observations and columns to covariates. We allow the dimensionality $p$ to be larger than the sample size $n$.  The unknown parameter vector
\(
\bbeta^*=(\beta_1^*,\dots,\beta_p^*)^\top\in\mathbb{R}^p
\)
characterizes the associations between the covariates and the response, while
\(
\bvarepsilon\sim\mathcal{N}(\mathbf{0},\sigma^2\mathbf{I}_n)
\)
denotes the random error vector, assumed to have mean zero and covariance
matrix \(\sigma^2\mathbf{I}_n\).
We define the {signal set} \( \cS \), often referred to as the active set, as the collection of indices where the entries of \( \bbeta^* \) are non-zero, and denote its complement by \( \cS^c \). 
Let \( \hat{\cS} \) be the set of indices corresponding to the features selected by a data-driven model selection procedure, such as LASSO \citep{tibshirani1996regression}, SCAD \citep{fan2001variable}, Elastic Net \citep{zou2005regularization}, or MCP \citep{zhang2010nearly}.
Our objective is to construct confidence intervals for the selected features indexed by \( \hat{\cS} \).

Before reviewing existing PoSI methods, we distinguish between two common inferential targets \citep{zhang2022post}. The first is the \emph{population-based target}, given by the selected components of the full-model coefficient vector,
\(
\bbeta^*_{\hat{\cS}}=\{\beta_j^*\}_{j\in\hat{\cS}}.
\)
The second is the \emph{projection-based target}. For any deterministic set $\mathcal M\subseteq\{1,\ldots,p\}$,
define
\begin{equation}
\label{def:proj_based}
\bbeta^\dagger_{\mathcal M}
=
\{\beta^\dagger_{j,\mathcal M}\}_{j\in\mathcal M}
=
\arg\min_{\bbeta_{\mathcal M}}
\left\|
\mathbb{E}[\by|\bX]-\bX_{\mathcal M}\bbeta_{\mathcal M}
\right\|_2^2,
\end{equation}
where $\bX_{\mathcal M}$ contains the columns of $\bX$ indexed
by $\mathcal M$. Under the full linear model, $\mathbb{E}[\by|\bX]=\bX\bbeta^*$. We use the superscript $\dagger$ to denote this
projection-based target, whose entries are referred to as
\textit{projection-based coefficients}. This target represents the best linear
approximation to the conditional mean $\bX\bbeta^*$ using
the features in $\mathcal M$.
In high-dimensional settings, inference for either target
requires additional conditions to overcome
the curse of dimensionality. These typically include the
sparsity assumption on $\bbeta^*$ and suitable conditions on $\bX$ \citep{zhao2006model} for
estimating $\bbeta^*_{\hat{\cS}}$, and full column rank of
$\bX_{\hat{\cS}}$ to ensure that
$\bbeta^\dagger_{\hat{\cS}}$ is uniquely defined. 



Beyond these identifiability considerations, these two targets answer different inferential questions.
The projection-based target $\bbeta^\dagger_{\hat{\cS}}$
depends on the selected features in the sub-model indexed by $\hat{\cS}$ and may absorb the effects of
correlated features omitted by the selection,
whereas each underlying coefficient $\beta_j^*$ is defined
independently of the selected sub-model. Consequently, the two
targets can differ substantially, with some components even
having opposite signs. When the full linear model is reasonably well specified, $\bbeta^*_{\hat{\cS}}$ may be more relevant for scientific interpretation. In contrast, interpreting inference for $\bbeta^\dagger_{\hat{\cS}}$ as inference for $\bbeta^*_{\hat{\cS}}$ can lead to misleading conclusions.

Existing PoSI methods have largely focused on projection-based targets. Valid PoSI seeks to construct confidence intervals for selected features that account for the bias induced by model selection while controlling an appropriate error criterion, such as simultaneous coverage, conditional coverage, or the FCR \citep{Kuchibhotla2021PostselectionI, zhang2022post}. To formalize these three criteria, let $\CI_{j,\mathcal{M}}$ denote the confidence interval for the coefficient of feature $j$
within a sub-model  containing features indexed by $\mathcal{M}$.

A pioneering contribution by \cite{berk2013valid} introduced the notion of
\textit{universal PoSI (UPoSI)}, which constructs CIs that are simultaneously
valid over all sub-models:
\[
\pr\left(
\beta^\dagger_{j,\mathcal{M}}\in\CI_{j,\mathcal{M}}
\text{ for all sub-models } \mathcal{M} \subseteq \{1, \dots, p\}
\text{ and all } j\in \mathcal{M}
\right)
\geq 1-\alpha.
\]
Consequently,
for any data-dependent selected sub-model $\hat{\cS}$,
\(
\pr\left(
\beta^\dagger_{j,\hat{\cS}}
\in
\CI_{j,\hat{\cS}}
\text{ for all }j\in\hat{\cS}
\right)
\geq 1-\alpha.
\)
Several works have proposed methodological extensions to improve upon the original UPoSI \citep{bachoc2019valid, bachoc2020uniformly, kuchibhotla2020valid, zhang2022post, Kuchibhotla2021PostselectionI}. Collectively, these proposals are referred to as UPoSI methods.


UPoSI methods can be computationally intensive and conservative
because they protect against arbitrary model selection,
often yielding wide confidence intervals
\citep{Kuchibhotla2021PostselectionI,zhang2022post}.
An alternative is \textit{exact post-selection inference}
(EPoSI), which conditions on the selection event
\citep{lee2016exact} and seeks to ensure
\[
\pr\left(
\beta^\dagger_{j,\mathcal M}\in\CI_{j,\mathcal M}
\,\middle|\,\hat{\cS}=\mathcal M
\right)\ge 1-\alpha,
\qquad j\in\mathcal M.
\]
Characterizing this conditional distribution generally requires
analysis specific to the selection procedure.
Extensions include other $\ell_1$-penalized models
\citep{taylor2018post}, reduced conditioning
\citep{liu2018more}, sequential regression procedures
\citep{tibshirani2016exact}, and change-point detection
\citep{hyun2021post}.


However, false coverage control alone does not ensure practically informative inference. A useful post-selection inference procedure should also retain sufficient power to identify genuinely associated features. Compared with simultaneous and conditional coverage guarantees used in UPoSI and EPoSI respectively, the FCR is a criterion that can better balance false coverage control and statistical power.
Due to this advantage, FCR has received growing attention in recent years \citep{xu2024post,FrostigBenjamini2025}, thereby motivating the development of new methods aiming to control FCR in PoSI. Other strategies separate the information used for selection
and inference. Data splitting partitions the observations
\citep{cox1975note}, while data fission and data thinning
decompose each observation into two observations used for these two tasks
\citep{leiner2025data,neufeld2024data}. Under suitable conditions,
these strategies can control the FCR, but may reduce power
by limiting the information available for inference.

The PoSI methods discussed above have largely focused on
projection-based targets, while methods for post-selection inference on $\bbeta^*_{\hat{\cS}}$ remain comparatively limited
\citep{zhang2022post}. The debiased LASSO
\citep{javanmard2014confidence,van2014asymptotically}
is a widely used approach to inference on the population-based
coefficient vector $\bbeta^*$ in high-dimensional linear models.
However, constructing the full debiasing matrix typically
requires solving $p$ additional optimization problems:
either nodewise LASSO regressions of each covariate on the
remaining $p-1$ covariates \citep{van2014asymptotically}, or constrained quadratic programs
that compute a separate debiasing direction for each coefficient \citep{javanmard2014confidence}. This computational burden can become prohibitive when $p$ is
large. Moreover, extending debiased-LASSO-based inference to account for data-dependent selection procedures beyond the LASSO generally requires procedure-specific theoretical analysis.


In this paper, we propose {TwinPoSI} (statistical twin post-selection inference), a confidence interval construction method that enables powerful post-selection inference for the population-based regression coefficient \(\bbeta^*_{\hat{\cS}}\) while controlling the FCR:
$$
\text{FCR} = \Ex \left( \frac{|\{j : \beta^*_{j} \notin \CI_{j,\hat{\cS}}, j \in \hat{\cS}\}|}{\max(|\hat{\cS}|,1)} \right).
$$
TwinPoSI uses \textit{statistical twins}, synthetic datasets generated from a fitted statistical model to mimic the original data. Although statistical twins are generated as parametric bootstrap samples, they are used differently from conventional bootstrap procedures: rather than approximating the sampling distribution of an estimator, they are used to calibrate the degree of expansion of initial confidence intervals to achieve FCR control.

The remainder of the paper is organized as follows. Section~\ref{sec:population-projection-targets} examines the relationship between the population-based and projection-based targets and identifies conditions under which their discrepancy vanishes asymptotically. Section~\ref{sec:twinposi} introduces TwinPoSI and its statistical-twin-based calibration procedure. Section~\ref{sec:theory} establishes asymptotic FCR control for TwinPoSI with respect to the population-based targets. In a Gaussian linear model with an orthogonal design, we further show that TwinPoSI asymptotically attains the oracle average interval length: its average interval length is asymptotically $1/\sqrt{2}$ times that of data splitting and data fission, while its length ratios relative to UPoSI and EPoSI converge to zero. Section~\ref{sec:simu} presents extensive simulation studies demonstrating reliable FCR control, shorter confidence intervals, and higher statistical power than existing methods across a wide range of settings, including high-dimensional regimes, model misspecification, and weak-signal scenarios. Finally, Section~\ref{sec:real} illustrates the practical utility of TwinPoSI in identifying genes associated with riboflavin production, while yielding the shortest average confidence intervals among the methods considered.

\section{Relationship Between Population-Based and Projection-Based Targets}
\label{sec:population-projection-targets}

TwinPoSI targets the selected population-based target
$\bbeta^*_{\hat{\cS}}$, whereas much of the existing PoSI
literature focuses on the projection-based target $\bbeta^\dagger_{\hat{\cS}}$.
For a fixed design matrix $\bX$ and any sub-model $\mathcal M$
such that $\bX_{\mathcal M}$ has full column rank,
\begin{equation}\label{eq:target_diff}
    \bbeta^\dagger_{\mathcal M}
=
\bbeta^*_{\mathcal M}
+
(\bX_{\mathcal M}^{\top}\bX_{\mathcal M})^{-1}
\bX_{\mathcal M}^{\top}\bX_{\mathcal M^c}
\bbeta^*_{\mathcal M^c},
\end{equation}
where $\mathcal M^c$ denotes the complement of $\mathcal M$.
The additional term captures the contribution of omitted
covariates and explicitly depends on the sub-model
$\mathcal M$. After selection, this relationship applies with
$\mathcal M=\hat{\cS}$ whenever
$\bX_{\hat{\cS}}$ has full column rank.

The projection-based target depends on the selected sub-model, so different selection procedures or even different tuning parameters can yield different projection-based targets for the same feature. This dependence affects both scientific interpretation and comparisons of false coverage and statistical power between the two targets. In particular, a selected feature with $\beta_j^*=0$ may nevertheless have
$\beta^\dagger_{j,\hat{\cS}}\ne0$ because of its correlation
with omitted signals; thus, a confidence interval for
$\beta^\dagger_{j,\hat{\cS}}$ that excludes zero therefore
does not necessarily imply a nonzero conditional association between feature $j$ and the response in the full model. Consequently, statistical power defined with respect to the population-based and projection-based targets is not directly comparable.

We examine this relationship from two complementary perspectives. First, we establish conditions under which the discrepancy between the two targets vanishes asymptotically, allowing the FCR guarantee of TwinPoSI to extend to projection-based targets. Second, we provide an explicit finite-sample example in which the two targets differ substantially, illustrating the consequences of interpreting one target as the other when their discrepancy is non-negligible.


\subsection{Asymptotic Agreement Under Sparse Riesz Condition}

We first study conditions under which the population-based and
projection-based targets are asymptotically close. Under suitable
conditions, mainly the sparse Riesz condition, on the design and the LASSO estimation error, we show that
\(
\left\|
\bbeta^\dagger_{\hat{\cS}}
-
\bbeta^*_{\hat{\cS}}
\right\|_2
=
o_{\pr}(1).
\)

\begin{theorem}
\label{thm:pop-proj}
Consider the linear model
$\by=\bX\bbeta^*+\bvarepsilon$, where
$\bvarepsilon\sim\mathcal{N}(\mathbf{0},\sigma^2\mathbf{I}_n)$.
Let $\hat{\cS}=\{j:\hat{\beta}_j\neq0\}$ be the active set selected by the
LASSO estimator $\hat{\bbeta}$, and let $s=\|\bbeta^*\|_0$. Suppose that
$s\log p/n\to0$. Assume that the design matrix $\bX$ satisfies the sparse
Riesz condition of \cite{Zhang2008Sparsity} and that the regularization
parameter $\lambda$ is chosen as in Theorem~3 of
\cite{Zhang2008Sparsity}. Then, with probability tending to one, the
projection-based coefficient $\bbeta^\dagger_{\hat{\cS}}$ is uniquely defined
and satisfies
$$\|\bbeta^\dagger_{\hat{\cS}}-\bbeta^*_{\hat{\cS}}\|_2
\leq C\|\hat{\bbeta}-\bbeta^*\|_2
=
O_{\pr}\!\left(\sigma\sqrt{s\log p/n}\right)
=
o_{\pr}(1).$$
\end{theorem}
Theorem~\ref{thm:pop-proj} shows that the discrepancy between
$\bbeta^*_{\hat{\cS}}$ and
$\bbeta^\dagger_{\hat{\cS}}$ vanishes when
$s\log p/n\to0$. 
This result does not imply that the two targets are interchangeable in
general, as it depends on both the sparse Riesz condition of the
design matrix and the consistency of the LASSO estimator. Combined with the additional boundary-stability conditions established in {Appendix}~D.4, this asymptotic agreement permits the FCR guarantee of TwinPoSI for the population-based target to be transferred to the projection-based target. 


\subsection{An Example Where the Two Targets Do Not Match}
The asymptotic agreement established in Theorem~\ref{thm:pop-proj} relies on regularity conditions that ensure a vanishing LASSO estimation error. The following example illustrates how, in finite samples, LASSO can omit numerous weak signals correlated with selected features, leading to a substantial discrepancy between the two targets and even a sign reversal. Consequently, interpreting valid inference for a projection-based coefficient as inference for its population-based counterpart can lead to misleading scientific conclusions.

\begin{example}
\label{ex:proj-mislead}
We consider a Gaussian linear model with \(n=1500\) observations, \(p=303\) features, and noise standard deviation \(\sigma=1.5\). The three strong features \(\bX_1,\bX_2,\bX_3\) form a correlated block with population-based coefficients \((\beta_1^*, \beta_2^*, \beta_3^*)^\top = (1,-1,1)^\top\), with pairwise correlations \(0.30\), \(0.92\), and \(0.60\) for \((\bX_1,\bX_2)\), \((\bX_1,\bX_3)\), and \((\bX_2,\bX_3)\), respectively. The next 100 weak features each have population-based coefficient \(0.04\) and correlations $0.06$, $-0.06$, and $0.06$ with $\bX_1$, $\bX_2$, and $\bX_3$, respectively. Their pairwise correlation is \(0.035\). The final 200 features are independent noise features with zero coefficients. The complete covariance structure is provided in Appendix~G.4.

Let \(\hat{\mathcal S}\) denote the model selected by LASSO. In one realization, \(\hat{\mathcal S}\) contains 45 features: \(\bX_1,\bX_2,\bX_3\), 39 of the 100 weak signals, and three noise features, leaving 61 weak signals unselected. Applying \eqref{eq:target_diff} to \(\hat{\mathcal S}\), the projection-based coefficient corresponding to \(\bX_1\) is
\[
\beta_{1,\hat{\mathcal S}}^\dagger
\approx -0.598<0,
\qquad\text{whereas}\qquad
\beta_1^*=1>0.
\]
Thus, despite the selection of many weak signals, the aggregate contribution of the omitted ones shifts the projection-based coefficient for \(\bX_1\) below zero. Consequently, a confidence interval lying entirely below zero can provide valid inference for \(\beta_{1,\hat{\mathcal S}}^\dagger\), even though it does not cover the population-based coefficient \(\beta_1^*>0\). Interpreting such an interval as evidence that the conditional association between \(\bX_1\) and the response given the other features is negative would therefore be misleading.
\end{example}

This example is further confirmed by simulation (Section~\ref{sec:target-mismatch-simu}), demonstrating that the target mismatch has direct inferential consequences: PoSI procedures targeting the projection-based coefficient predominantly produce confidence intervals for the coefficient of \(\bX_1\) that either lie entirely below zero or contain zero, whereas TwinPoSI, which targets the population-based coefficient, predominantly produces intervals lying entirely above zero.

\section{The TwinPoSI Methodology}
\label{sec:twinposi}
\subsection{Overview of TwinPoSI} 
TwinPoSI constructs post-selection confidence intervals by calibrating their expansion using synthetic datasets, which we refer to as \emph{statistical twins}. A statistical twin is generated from a fitted model to mimic the original data-generating mechanism, but, unlike the original data, its generating parameters are known. This known-truth setting allows us to evaluate post-selection CI coverage under the same estimation and selection procedure and thereby calibrate the correction factor needed to control the FCR.

Let
\[
\mathcal{E}(\cdot,\cdot):
\mathbb{R}^n\times\mathbb{R}^{n\times p}\to\mathbb{R}^p
\]
denote a generic estimation and feature-selection procedure that maps a response vector and a design matrix to an estimated coefficient vector, such as the LASSO, SCAD, or Elastic Net. We primarily focus on the LASSO and construct CIs for the selected population-based coefficients. We further demonstrate empirically that TwinPoSI controls the FCR while yielding short confidence intervals under SCAD and Elastic Net selection (Section~\ref{sec:selection-methods}). Applying \(\mathcal{E}\) to the original data \((\by,\bX)\) yields the estimated coefficient vector and the corresponding selected set:
\begin{equation}\label{eq:beta_hat}
\hat{\bbeta}
=
\mathcal{E}(\by,\bX),
\qquad
\hat{\mathcal S}
=
\{j:\hat\beta_j\neq0\}.
\end{equation}

Suppose that, for each feature \(j\in\{1,\ldots,p\}\), we construct an initial CI
\[
\hat I_j^{\mathrm{init}}
=
[\hat c_j^{\mathrm{init}},\hat C_j^{\mathrm{init}}]
\]
at the nominal level \(1-q\) using a conventional interval-construction procedure, such as the bootstrap. In our implementation, these are basic bootstrap CIs obtained from the empirical distribution of the feature-wise estimation errors, as detailed in Section~\ref{sec:procedure}. The bootstrap captures the sampling variability of each estimated coefficient under the estimation procedure \(\mathcal E\), but the resulting CIs are constructed feature-wise without calibrating for the fact that only features selected from the same data are subsequently reported. These initial CIs are constructed for all features to facilitate the calibration in Section~\ref{sec:procedure}, although only the calibrated CIs for the selected features \(j\in\hat{\cS}\) are ultimately reported. Consequently, the nominal coverage of the initial CIs does not in general translate into FCR control after data-dependent feature selection; in particular, there is no guarantee that
\[
\mathbb{E}\left[
\frac{
|\{j\in\hat{\mathcal S}:\beta_j^*\notin\hat I_j^{\mathrm{init}}\}|
}{
|\hat{\mathcal S}|\vee1
}
\right]
\le q.
\]

To account for the selection-induced loss of coverage, we consider \emph{calibrated CIs} obtained by symmetrically expanding each initial CI by an amount \(t\geq 0\):
\[
\hat I_j(t)
=
[\hat c_j^{\mathrm{init}}-t,\ \hat C_j^{\mathrm{init}}+t],
\qquad
j\in\hat{\mathcal S}.
\]
For a target FCR level \(q\), define the \emph{oracle correction factor}
\begin{equation}
\label{eq:oracle-correction}
T_q
=
\mathcal L_q(\bbeta^*,\sigma^2)
:=
\inf\left\{
t\geq0:
\mathbb E\left[
\frac{
|\{j\in\hat{\mathcal S}:\beta_j^*\notin\hat I_j(t)\}|
}{
|\hat{\mathcal S}|\vee1
}
\right]
\leq q
\right\}.
\end{equation}
Here, the expectation is taken over the data-generating distribution under the true parameters \((\bbeta^*,\sigma^2)\). Thus, \(T_q\) represents the smallest additive correction factor required to reduce the post-selection FCR to at most \(q\). The correction factor is well defined because the FCR is non-increasing in \(t\) and converges to zero as \(t\to\infty\). If \((\bbeta^*,\sigma^2)\) were known, the resulting oracle-calibrated CIs would be
\[
\hat I_j(T_q)
=
[\hat c_j^{\mathrm{init}}-T_q,\ \hat C_j^{\mathrm{init}}+T_q],
\qquad
j\in\hat{\mathcal S}.
\]
In practice, \((\bbeta^*,\sigma^2)\) is unknown, so \(T_q\) cannot be evaluated directly. We next introduce statistical twins to estimate this oracle correction factor.

\subsection{Statistical Twins}

Because the oracle correction factor \(T_q=\mathcal{L}_q(\bbeta^*,\sigma^2)\) depends on the unknown parameters \((\bbeta^*,\sigma^2)\), it cannot be evaluated directly. Motivated by the plug-in strategy of Nullstrap \citep{wang2025nullstrapsimplehighpowerfast}, we use
\[
\hat{\bbeta}=\mathcal{E}(\by,\bX)
\]
as an estimate of \(\bbeta^*\) and estimate \(\sigma^2\) by a consistent variance estimator \(\hat{\sigma}^2\) constructed from the scaled residuals \citep{reid2016study}. We then approximate the original data-generating model by the fitted model with parameters \((\hat{\bbeta},\hat{\sigma}^2)\). This motivates the following parametric construction of a statistical twin.

\begin{definition}[Parametric statistical twin]\label{def:statistical_twin}
For the linear model
\[
\by=\bX\bbeta^*+\bvarepsilon,
\qquad
\bvarepsilon\sim\mathcal{N}(\mathbf 0,\sigma^2\mathbf I_n),
\]
let \((\hat{\bbeta},\hat{\sigma}^2)\) be estimates of \((\bbeta^*,\sigma^2)\) obtained from the original data \((\by,\bX)\). A \emph{parametric statistical twin} is a synthetic response vector \(\tilde{\by}\in\mathbb{R}^n\) generated as
\[
\tilde{\by}
=
\bX\hat{\bbeta}
+
\hat\sigma\bZ,
\qquad
\bZ\sim\mathcal{N}(\mathbf 0,\mathbf I_n),
\]
where \(\bZ\) is independent of the original data. Conditional on the original data,
\[
\tilde{\by}
\sim
\mathcal{N}
\left(
\bX\hat{\bbeta},
\hat\sigma^2\mathbf I_n
\right).
\]
\end{definition}

The data-generation mechanism in Definition~\ref{def:statistical_twin} is the same as that of a parametric bootstrap sample, but its inferential role in TwinPoSI is different. Rather than using the synthetic datasets primarily to approximate the sampling distribution of an estimator, TwinPoSI exploits their known generating parameters to evaluate post-selection CI coverage under the same estimation and selection procedure as applied to the original data. Conditional on the original data, \((\hat{\bbeta},\hat{\sigma}^2)\) are fixed and therefore constitute the known generating parameters in the statistical-twin world. In particular, \(\hat{\bbeta}\) plays the role of the known population-based coefficient vector, allowing post-selection coverage to be evaluated directly and thereby providing the basis for estimating the correction factor required for FCR control.

A nonparametric version of the statistical twin, which accommodates non-Gaussian errors by resampling centered and scaled residuals, is provided in Appendix~B.1.


\subsection{TwinPoSI Procedure}\label{sec:procedure}

TwinPoSI proceeds in two stages: construction of the initial CIs and calibration of their expansion. The first stage uses parametric bootstrap samples to estimate feature-wise uncertainty, whereas the second stage uses an independent set of statistical twins to calibrate the additional uncertainty induced by data-dependent feature selection. All fits use the same estimation and feature-selection procedure \(\mathcal E\) as that applied to the original data. For the LASSO, the tuning parameter is held fixed at the value used for the original fit.

We first construct the initial CIs. Starting from
\(\hat{\bbeta}=\mathcal E(\by,\bX)\) and \(\hat{\sigma}^2\), we generate \(R\) parametric bootstrap samples, with \(R=20\) by default:
\[
\by^{(r)}
=
\bX\hat{\bbeta}
+
\hat{\sigma}\mathbf Z^{(r)},
\qquad
\hat{\bbeta}^{(r)}
=
\mathcal E(\by^{(r)},\bX),
\qquad
r=1,\ldots,R,
\]
where the \(\mathbf Z^{(r)}\) are independent standard Gaussian vectors. For each feature \(j=1,\ldots,p\), define the bootstrap deviations
\[
d_j^{(r)}
=
\hat\beta_j-\hat\beta_j^{(r)},
\qquad
r=1,\ldots,R.
\]
The initial basic bootstrap CI
\[
\hat I_j^{\mathrm{init}}
=
[\hat c_j^{\mathrm{init}},\hat C_j^{\mathrm{init}}]
\]
is constructed from the empirical quantiles of these deviations:
\begin{equation}\label{eq:initial_endpoints}
\begin{aligned}
\hat c_j^{\mathrm{init}}
&=
\hat\beta_j+
\operatorname{Quantile}_{q/2}
\left(
\{d_j^{(r)}\}_{r=1}^R
\right),\\
\hat C_j^{\mathrm{init}}
&=
\hat\beta_j+
\operatorname{Quantile}_{1-q/2}
\left(
\{d_j^{(r)}\}_{r=1}^R
\right).
\end{aligned}
\end{equation}
These initial CIs are constructed for all \(p\) features because a feature not selected from the original data may be selected in a calibration statistical twin, in which case its construction-stage interval offsets are needed. As discussed in Section~\ref{sec:twinposi}, these feature-wise CIs are not yet calibrated for data-dependent feature selection.

We next calibrate the interval expansion using an independent set of \(B\) statistical twins, with \(B=B_n=\lceil\log n\rceil\) by default. Generate
\[
\tilde{\by}^{(b)}
=
\bX\hat{\bbeta}
+
\hat{\sigma}\tilde{\mathbf Z}^{(b)},
\qquad
b=1,\ldots,B,
\]
where the \(\tilde{\mathbf Z}^{(b)}\) are independent standard Gaussian vectors and are independent of both the original data and the bootstrap samples used to construct the initial CIs. Applying the same procedure \(\mathcal E\) to each statistical twin gives
\begin{equation}\label{eq:calibration_fit}
\tilde{\bbeta}^{(b)}
=
\mathcal E(\tilde{\by}^{(b)},\bX),
\qquad
\tilde{\cS}^{(b)}
=
\{j:\tilde\beta_j^{(b)}\ne0\}.
\end{equation}

To construct the corresponding initial CIs in each statistical twin without introducing an additional layer of bootstrap sampling, we reuse the feature-wise lower and upper offsets obtained from the first stage. Specifically, for each \(j\in\tilde{\cS}^{(b)}\), define
\begin{equation}\label{eq:twin_endpoints}
\begin{aligned}
\tilde c_j^{\mathrm{init},(b)}
&=
\tilde\beta_j^{(b)}
+
(\hat c_j^{\mathrm{init}}-\hat\beta_j),\\
\tilde C_j^{\mathrm{init},(b)}
&=
\tilde\beta_j^{(b)}
+
(\hat C_j^{\mathrm{init}}-\hat\beta_j).
\end{aligned}
\end{equation}
Thus, the construction-stage offsets are computed once and shared across all calibration statistical twins. For an expansion amount \(t\ge0\), define
\[
\tilde I_j^{(b)}(t)
=
[
\tilde c_j^{\mathrm{init},(b)}-t,\,
\tilde C_j^{\mathrm{init},(b)}+t
],
\qquad
j\in\tilde{\cS}^{(b)}.
\]

Because \(\hat{\bbeta}\) is the known generating coefficient vector in the statistical-twin world, we can directly determine whether the CI for each selected feature covers its target coefficient. We estimate the twin-world FCR at expansion amount \(t\) by
\begin{equation}\label{eq:basic}
\widehat{\operatorname{FCR}}_{\mathrm{twin}}(t)
=
\frac{1}{B}
\sum_{b=1}^B
\frac{
\left|
\left\{
j\in\tilde{\cS}^{(b)}:
\hat\beta_j\notin\tilde I_j^{(b)}(t)
\right\}
\right|
}{
|\tilde{\cS}^{(b)}|\vee1
}.
\end{equation}
We then choose the smallest expansion amount for which the estimated twin-world FCR does not exceed \(q\):
\begin{equation}\label{eq:calibration_threshold}
\hat T_q
=
\inf\left\{
t\ge0:
\widehat{\operatorname{FCR}}_{\mathrm{twin}}(t)
\le q
\right\}.
\end{equation}
Thus, \(\hat T_q\) is a Monte Carlo estimate of the correction factor in the statistical-twin world, \(\mathcal L_q(\hat{\bbeta},\hat{\sigma}^2)\), which serves as a plug-in approximation to the oracle correction factor \(T_q=\mathcal L_q(\bbeta^*,\sigma^2)\).

Finally, the TwinPoSI CIs for the features selected from the original data are
\begin{equation}\label{eq:final_interval}
\mathrm{CI}_{j,\hat{\cS}}
=
\hat I_j(\hat T_q)
=
[
\hat c_j^{\mathrm{init}}-\hat T_q,\,
\hat C_j^{\mathrm{init}}+\hat T_q
],
\qquad
j\in\hat{\cS}.
\end{equation}

Appendix~B.2 provides a more conservative calibration rule that takes the maximum of the expansion amounts required to attain a noncoverage proportion at or below \(q\) in each calibration statistical twin. Sensitivity to the number of calibration statistical twins is examined in Appendix~H.

\begin{algorithm}[t]
\caption{TwinPoSI}
\label{alg:TwinPoSI_final}
\small
\KwData{
Response $\by$, design matrix $\bX$, estimation and feature-selection procedure
$\mathcal E$, target FCR level $q$, number of parametric bootstrap samples $R$, and number of calibration statistical twins $B$.
}
\KwResult{
Selected set $\hat{\cS}$ and calibrated CIs
$\{\mathrm{CI}_{j,\hat{\cS}}:j\in\hat{\cS}\}$.
}

Compute $\hat{\bbeta}=\mathcal E(\by,\bX)$,
estimate $\hat{\sigma}^2$, and set
$\hat{\cS}=\{j:\hat\beta_j\ne0\}$\;

\textbf{Step 1: Construct initial CIs using parametric bootstrap samples}\;
\For{$r=1,\ldots,R$}{
Draw $\mathbf Z^{(r)}
\sim\mathcal N(\mathbf0,\mathbf I_n)$\;
Generate the bootstrap sample
$\by^{(r)}
=\bX\hat{\bbeta}+\hat{\sigma}\mathbf Z^{(r)}$\;
Compute
$\hat{\bbeta}^{(r)}=\mathcal E(\by^{(r)},\bX)$\;
}
For each $j=1,\ldots,p$, compute
$d_j^{(r)}=\hat\beta_j-\hat\beta_j^{(r)}$, $r=1,\ldots,R$,
and obtain $\hat c_j^{\mathrm{init}}$ and
$\hat C_j^{\mathrm{init}}$ using \eqref{eq:initial_endpoints}\;

\textbf{Step 2: Calibrate CI expansion using statistical twins}\;
\For{$b=1,\ldots,B$}{
Draw $\tilde{\mathbf Z}^{(b)}
\sim\mathcal N(\mathbf0,\mathbf I_n)$\;
Generate the calibration statistical twin
$\tilde{\by}^{(b)}
=\bX\hat{\bbeta}+\hat{\sigma}\tilde{\mathbf Z}^{(b)}$\;
Compute
$\tilde{\bbeta}^{(b)}
=\mathcal E(\tilde{\by}^{(b)},\bX)$
and set
$\tilde{\cS}^{(b)}
=\{j:\tilde\beta_j^{(b)}\ne0\}$\;
For each $j\in\tilde{\cS}^{(b)}$, set
$\tilde c_j^{\mathrm{init},(b)}
=\tilde\beta_j^{(b)}+(\hat c_j^{\mathrm{init}}-\hat\beta_j)$
and
$\tilde C_j^{\mathrm{init},(b)}
=\tilde\beta_j^{(b)}+(\hat C_j^{\mathrm{init}}-\hat\beta_j)$\;
Define
$\tilde I_j^{(b)}(t)
=
[\tilde c_j^{\mathrm{init},(b)}-t,
\tilde C_j^{\mathrm{init},(b)}+t]$,
$t\ge0$\;
}

Compute $\widehat{\operatorname{FCR}}_{\mathrm{twin}}(t)$
using \eqref{eq:basic} and obtain $\hat T_q$
from \eqref{eq:calibration_threshold}\;

\Return{
$\hat{\cS}$ \textnormal{and}
$\{[\hat c_j^{\mathrm{init}}-\hat T_q,
\hat C_j^{\mathrm{init}}+\hat T_q]:
j\in\hat{\cS}\}$
}
\end{algorithm}

\section{Theoretical Properties of TwinPoSI}
\label{sec:theory}

\subsection{Asymptotic FCR Control}
We establish FCR control through a comparison of the noncoverage
proportions in the original and statistical-twin experiments.
Let $\mathcal C$ denote the information generated by the independent Gaussian noise vectors used in the initial interval construction, and let
$\mathcal G\supseteq\mathcal C$ contain all pre-calibration information,
including the original data, fitted quantities, and initial interval
construction. Precise definitions of $\mathcal C$ and $\mathcal G$ are provided
in the Appendix~D. The interval endpoints may depend on the original data.
Conditional on $\mathcal G$, the $B$ calibration twins are independent
and identically distributed.

For $t\ge0$, define
\begin{equation*}
\mathcal{R}_{\mathrm{orig}}(t)
=
\mathbb E\left[
\left.
\frac{\bigl|\{j\in\hat \cS:\beta_j^*\notin\hat I_j(t)\}\bigr|}
{|\hat \cS|\vee1}
\right|\mathcal C
\right],
\qquad
\mathcal{R}_{\mathrm{twin}}(t)
=
\mathbb E\left[
\left.
\frac{\bigl|\{j\in\tilde \cS:\hat\beta_j\notin\tilde I_j(t)\}\bigr|}
{|\tilde \cS|\vee1}
\right|\mathcal G
\right].
\label{eq:phi}
\end{equation*}
The expectation defining $\mathcal R_{\mathrm{twin}}(t)$ is taken over the randomness of a calibration twin, conditional on $\mathcal G$. In practice, this conditional risk is estimated by averaging over the $B$ calibration twins. Fractions corresponding to empty selected sets are defined as zero. These functions describe the expected proportions of
reported intervals that fail to cover their respective targets
at correction level $t$.

\begin{assumption}[Uniform conditional-tail domination]
\label{ass:tail-domination}
As $n,p\to\infty$,
\begin{equation*}
\delta_n
:=
\sup_{t>0}
\bigl[
\mathcal{R}_{\mathrm{orig}}(t)
-
\mathcal{R}_{\mathrm{twin}}(t)
\bigr]_+
\xrightarrow{P}0,
\qquad [x]_+=\max\{x,0\}.
\label{eq:tail-domination}
\end{equation*}
\end{assumption}

Assumption~\ref{ass:tail-domination} requires that statistical twins
do not systematically underestimate the original noncoverage
proportion across correction levels. The comparison is one-sided:
overestimation is permitted. It does not require equality of the
original and twin selected sets or consistent recovery of the true
support. Appendix~E provides sufficient conditions for
this assumption under LASSO with a correlated fixed
design. These
conditions require stability on a deterministic reference set,
but do not require exact recovery of the true support.

\begin{theorem}
\label{thm:twinposi-fcr}
Suppose the initial interval endpoints are finite, and the
calibration twins are conditionally independent and identically
distributed given $\mathcal G$. For the empirical-average
calibration rule defining $\hat T_q$, every $\tau>0$ satisfies
\begin{equation*}
\operatorname{FCR}
\le
q+\tau+\pr(\delta_n>\tau)
+\sqrt{\frac{\pi}{8B}}.
\label{eq:fcr-finite}
\end{equation*}
Under Assumption~\ref{ass:tail-domination}, the reported intervals
satisfy
\begin{equation*}
\mathbb E\left[
\frac{
\bigl|\{j\in\hat \cS:
\beta_j^*\notin\hat I_j(\hat T_q)\}\bigr|
}{
|\hat \cS|\vee1
}
\right]
\le
q+\sqrt{\frac{\pi}{8B}}+o(1).
\label{eq:fcr-asymptotic}
\end{equation*}
The negligible term tends to zero as $n,p\to\infty$ and is
independent of the calibration count $B\ge1$. In particular, if $B=B_n\to\infty$, then $\operatorname{FCR}\le q+o(1)$, yielding asymptotic FCR control at the nominal level $q$.
\end{theorem}

Theorem~\ref{thm:twinposi-fcr} separates the error from using
statistical twins to approximate the original experiment from
the Monte Carlo error due to finitely many calibration replicates.
Specifically, $\delta_n$ measures the largest underestimation
of the original conditional noncoverage proportion by the twin
experiment across correction levels. For any $\tau>0$, on the event $\{\delta_n\le\tau\}$, the twin
risk curve underestimates the original risk curve by at most $\tau$.
The term $\pr(\delta_n>\tau)$ accounts for the complementary event.
Under Assumption~\ref{ass:tail-domination}, these approximation
terms can jointly be made asymptotically negligible, whereas
the Monte Carlo allowance decreases at rate $B^{-1/2}$.
Consequently, $B_n\to\infty$ yields
$\operatorname{FCR}\le q+o(1)$.
The proof is provided in the
Appendix~D.

\subsection{Asymptotic Efficiency of Confidence Intervals}
\label{sec:ci-efficiency-example}

We next examine the interval-length efficiency of TwinPoSI under a Gaussian linear model with an orthogonal design.
Orthogonality ensures that the
population-based and projection-based coefficient targets coincide.  
For each procedure $A$, let $\ell^A$ denote the arithmetic mean length of all intervals reported for its selected features. Consider the Gaussian linear model with known $\sigma>0$, fixed
$q\in(0,1)$, and
\begin{equation*}
 n=2p,\qquad
 \mathbf X=\sqrt p\begin{pmatrix}\mathbf I_p\\ \mathbf I_p\end{pmatrix},\qquad
 \frac{\mathbf X^\top\mathbf X}{n}=\mathbf I_p.
 \label{eq:example-design}
\end{equation*}
Let $s=\lfloor p^{1/2}\rfloor$, $\mathcal S=\{1,\ldots,s\}$, and set
\begin{equation*}
 \lambda=\frac{\sigma}{\sqrt n}\,
   \Phi^{-1}\!\left(1-\frac{s}{2p(\log p)^{1/4}}\right),\qquad
 \beta_j^*=4\lambda\,\mathbf1\{j\in\mathcal S\}.
 \label{eq:example-sequence}
\end{equation*}
Use LASSO with penalty $\lambda$ and apply TwinPoSI with
$\hat\sigma=\sigma$, $R=\lceil(\log p)^2\rceil$, and any deterministic
integer $1\le B\le R$. Here, $R$ and $B$ denote the numbers of construction and calibration twins, respectively.
 Data splitting (DS) uses the two blocks of $\mathbf X$ for half-sample splitting, and data fission (DF)
uses equal-noise Gaussian fission. UPoSI and EPoSI use the original LASSO
selected set. 
Precise
procedure definitions are given in the Appendix~F.

For comparison, an oracle knows the true support $\mathcal S$ and $\sigma$
and fits the true submodel by least squares using all $n$ observations. Since
$\mathbf X_{\mathcal S}^\top\mathbf X_{\mathcal S}=n\mathbf I_s$, its
$(1-q)$-intervals for $j\in\mathcal S$ and their average length are
\[
 \left[\frac{\mathbf X_j^\top\mathbf y}{n}
       \ \pm\ z_{1-q/2}\frac{\sigma}{\sqrt n}\right],\qquad
 \ell^{\mathrm{oracle}}=2z_{1-q/2}\frac{\sigma}{\sqrt n},
\]
where $\mathbf X_j$ is column $j$ of $\mathbf X$ and
$z_{1-q/2}=\Phi^{-1}(1-q/2)$.

\begin{theorem}
\label{thm:orthogonal-length-comparison}
As $n,p\to\infty$,
\begin{equation}
 \begin{aligned}
 \frac{\ell^{\mathrm{TwinPoSI}}}{\ell^{\mathrm{oracle}}}
       &\xrightarrow{\mathbb P}1,\\
 \frac{\ell^{\mathrm{TwinPoSI}}}{\ell^{\mathrm{DS}}},\quad
 \frac{\ell^{\mathrm{TwinPoSI}}}{\ell^{\mathrm{DF}}}
       &\xrightarrow{\mathbb P}\frac1{\sqrt2},\\
 \frac{\ell^{\mathrm{TwinPoSI}}}{\ell^{\mathrm{UPoSI}}},\quad
 \frac{\ell^{\mathrm{TwinPoSI}}}{\ell^{\mathrm{EPoSI}}}
       &\xrightarrow{\mathbb P}0.
 \end{aligned}
 \label{eq:example-length-ratios}
\end{equation}
\end{theorem}

The first limit shows that TwinPoSI attains the oracle average interval length to first order. Its average length is asymptotically $1/\sqrt{2}$ times that of the specified DS and DF procedures. In this example, DS uses half the observations for inference, while equal-noise DF doubles the inference noise variance; both retain half the full-data information for inference, inflating standard errors by $\sqrt{2}$. TwinPoSI also has average interval length of smaller order than those of UPoSI and EPoSI.  The proof is given in Appendix F.

\section{Simulation Studies}
\label{sec:simu}
\subsection{Comparison of Eight PoSI Methods}

We compare TwinPoSI with seven other approaches in a low-dimensional setting where all eight methods are computationally feasible: UPoSI \citep{kuchibhotla2020valid}, EPoSI \citep{lee2016exact}, EPoSI for forward selection (EPoSI-FS), EPoSI for least angle regression (EPoSI-LARS) \citep{tibshirani2016exact}, data fission, data splitting, and Benjamini--Yekutieli-adjusted debiased LASSO (BY-dLASSO).

Throughout the simulation studies, we evaluate FCR with respect to the population-based target $\bbeta^*_{\hat{\cS}}$, consistent with the primary inferential objective of TwinPoSI.
We also report $\operatorname{FCR}^{\dagger}$, the false coverage rate evaluated with respect to the projection-based target $\bbeta^\dagger_{\hat{\cS}}$.
For methods originally developed for projection-based targets, the reported FCR therefore evaluates whether their confidence intervals also cover the
corresponding population-based coefficients.
The CI length measures inferential precision and is defined as the average
width of the confidence intervals over the selected set $\hat{\cS}$. Shorter
intervals, subject to FCR control at the target level, indicate greater
inferential precision. We further define efficient power as
\[
\operatorname{Efficient\ Power}
=
\Ex\left[
\frac{
\left|
\left\{
j\in\cS\cap\hat{\cS}:
\beta_j^*\in\CI_{j,\hat{\cS}},
\ 0\notin\CI_{j,\hat{\cS}}
\right\}
\right|
}{
\max(|\cS|,1)
}
\right].
\]
This metric quantifies the proportion of true signals that are selected,
covered by their confidence intervals, and declared statistically significant.

We first consider the following low-dimensional simulation setting, where the number of features is smaller than the sample size.
\begin{simsetting}\label{setting:1}

We set $n=300$ and $p \in \{50, 100\}$. The design matrix $\mathbf{X}$ consists of i.i.d. rows and AR(1) columns, generated from $\mathcal{N}(\bold{0},\bSigma)$, where $\bSigma$ is a Toeplitz matrix with an autocorrelation parameter $\rho$, representing the correlation between two adjacent variables in $\mathbf{X}$. 
Specifically, $\rho$ is set to $0.5$ in this setting. The first 5 elements of the coefficient $\bbeta^*$ are assigned values with amplitude $A = \frac{\log{p}}{(1-|\rho|)\sqrt{n}} $ and random signs while the other entries are set to zero. The response vector follows $\mathbf{y} = \mathbf{X} \bbeta^* + \bvarepsilon$, where $\bvarepsilon \sim \mathcal{N}(\bold{0}, \mathbf{I}_n)$. The target FCR level is set to $0.1$ for all methods. Each method is repeated 100 times.
\end{simsetting}


Due to the computational intensity of the original UPoSI procedure, we adopt the computationally efficient UPoSI implementation proposed by \cite{kuchibhotla2020valid}.
 The EPoSI, EPoSI-FS and EPoSI-LARS methods are implemented using the R package \texttt{selectiveInference}; they perform selective inference by conditioning on the model selected by LASSO, forward selection, and least angle regression, respectively. For data splitting and data fission, we follow the implementation details described in \cite{leiner2025data}.  Additionally, we implement the debiased LASSO using the R package \texttt{hdi} \citep{Dezeure2015hdi} and combine it with the Benjamini–Yekutieli procedure (BY-dLASSO), which provides valid PoSI by constructing FCR-adjusted confidence intervals.

The simulation results, summarized in Table~\ref{tab:setting1}, show that TwinPoSI consistently controls the FCR at the target level while attaining the highest efficient power and the shortest CI length among all methods. Data splitting and data fission often fail to control the FCR at level $q$ when evaluated with respect to the population-based target, and they yield longer CIs with substantially lower efficient power than TwinPoSI. 
UPoSI is notably conservative and produces excessively wide CIs and a consequent loss in statistical power. {The EPoSI methods (EPoSI, EPoSI-FS, and EPoSI-LARS) can yield very wide confidence intervals \citep{kivaranovic2021on}; their interval lengths also exhibit substantial variability across repetitions.} BY-dLASSO ranks second in terms of efficient power, but its FCR remains slightly inflated. In addition, Table~S1 in {Appendix}~G.1 compares the average runtime of TwinPoSI with those of other methods. The results show that all methods, except BY-dLASSO, are computationally efficient. 



\begin{table}[htbp]
\centering
\caption{Comparison of methods under Simulation Setting~\ref{setting:1}. Results are averaged over 100 repetitions. The target FCR level is $0.1$. Bold values highlight favorable results.}
\label{tab:setting1}
\small
\setlength{\tabcolsep}{3.5pt}
\begin{tabular}{lcccc@{\hspace{1em}}cccc}
\toprule
& \multicolumn{4}{c}{$p=50$} & \multicolumn{4}{c}{$p=100$} \\
\cmidrule(lr){2-5} \cmidrule(lr){6-9}
Method & Eff. Power & FCR & FCR$^\dagger$ & CI Length
       & Eff. Power & FCR & FCR$^\dagger$ & CI Length \\
\midrule
TwinPoSI
& \textbf{0.978} & \textbf{0.048} & \textbf{0.048} & \textbf{0.262}
& \textbf{0.972} & \textbf{0.052} & \textbf{0.052} & \textbf{0.231} \\
Data Fission
& 0.832 & 0.124 & 0.102 & 0.323
& 0.880 & 0.108 & \textbf{0.100} & 0.319 \\
Data Splitting
& 0.834 & 0.117 & \textbf{0.084} & 0.325
& 0.804 & 0.134 & 0.122 & 0.316 \\
UPoSI
& 0.032 & \textbf{0.000} & \textbf{0.000} & 1.331
& 0.022 & \textbf{0.000} & \textbf{0.000} & 1.588 \\
EPoSI
& 0.828 & 0.111 & \textbf{0.072} & 1.087
& 0.812 & 0.125 & \textbf{0.100} & 1.119 \\
EPoSI-FS   
& 0.474 & \textbf{0.073} & \textbf{0.073} & 5.768
           & 0.296 & \textbf{0.093} & \textbf{0.093} & 16.941 \\
EPoSI-LARS & 0.178 & 0.118 & 0.118 & 27.484
           & 0.076 & \textbf{0.091} & \textbf{0.091} & 110.835 \\
BY-dLASSO
& 0.956 & 0.108 & 0.108 & 0.351
& 0.924 & 0.102 & 0.102 & 0.374 \\
\bottomrule
\end{tabular}
\end{table}
\subsection{Further Comparison of Five PoSI Methods}

We next compare TwinPoSI with data splitting, data fission, UPoSI, and EPoSI in high-dimensional settings, evaluating FCR, $\operatorname{FCR}^{\dagger}$, CI length, and efficient power. We exclude EPoSI-FS and EPoSI-LARS because of their wide intervals in the preceding comparison (Table~\ref{tab:setting1}), and BY-dLASSO because of its high computational cost. For visual clarity, the main figures display only TwinPoSI, data splitting, and data fission, as the substantially wider UPoSI and EPoSI intervals would obscure differences among these methods. Full numerical results for all five methods are provided in Appendix~G.2.


\begin{simsetting}
\label{setting:2}
    We set $n = 1000$. The first 20 elements of the coefficient $\bbeta^*$ are assigned values with amplitude $A $ and random signs while the other entries are set to zero. The response vector $\mathbf{y}$ and design matrix $\mathbf{X} $ are generated as in Simulation Setting~\ref{setting:1}. We vary one of the following three parameters at a time: (a) the number of covariates $p \in \{1000,2000,3000,4000\}$; (b) the autocorrelation parameter $\rho \in [0, 0.7]$; (c) the target FCR level $q \in [0.05,0.4]$. For each scenario where one simulation parameter varies, the remaining parameters are held constant as:
    $$
    \rho = 0.5, q=0.1, p=1000, A= \frac{\log{p}}{(1-|\rho|)\sqrt{n}}.
    $$
\end{simsetting}

\begin{figure}[htbp]
    \centering
    \includegraphics[width=1\linewidth]{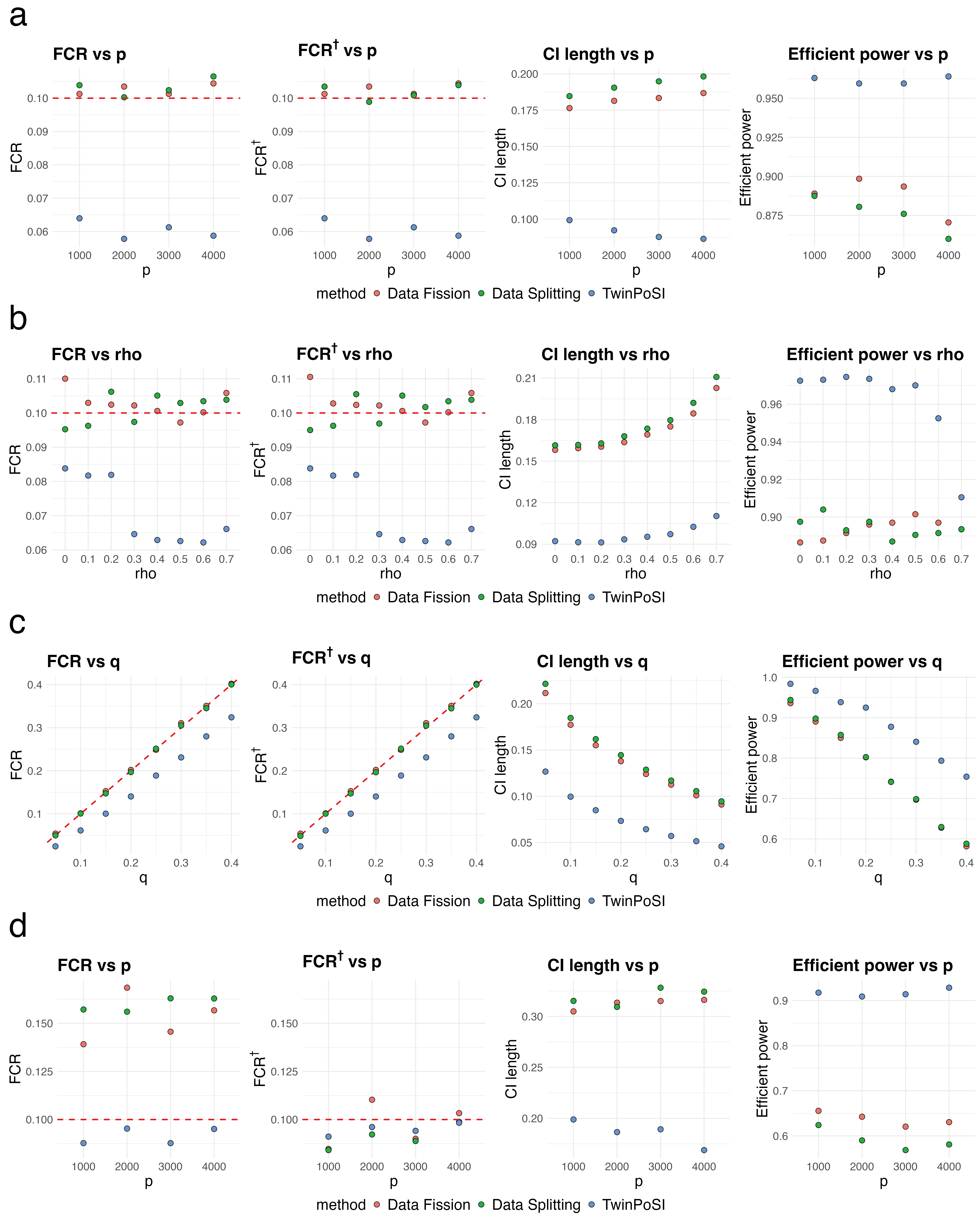}
   \caption{FCR, $\operatorname{FCR}^{\dagger}$, CI length, and efficient power across simulation settings.
FCR and $\operatorname{FCR}^{\dagger}$ are evaluated with
respect to the population-based target $\bbeta^*_{\hat{\cS}}$
and the projection-based target $\bbeta^\dagger_{\hat{\cS}}$,
respectively.
Panels (a)--(c) correspond to Simulation
Setting~\ref{setting:2}(a)--(c), varying $p$, $\rho$, and $q$,
respectively. Panel (d) varies $p$ under Student's
$t$-distributed errors with 3 degrees of freedom,
as specified in Simulation Setting~S1 in Appendix~G.3. In panels (a)--(c), the similar FCR and $\operatorname{FCR}^{\dagger}$ values reflect the strong-signal settings, in which the selected models typically contain the true support and the two targets therefore coincide.}
\label{fig:settings2-3}
\end{figure}

For each scenario under Simulation Setting~\ref{setting:2}, we evaluate TwinPoSI alongside four competing methods based on 100 independent simulation replicates.  
The performances of TwinPoSI, data fission, and data splitting are displayed in Figure~\ref{fig:settings2-3}.  
Across the scenarios considered, TwinPoSI maintains empirical
population-based FCR near or below the target level and achieves higher
efficient power and shorter average confidence intervals than the competing
methods shown in Figure~1(a)--(c). Its empirical $\operatorname{FCR}^{\dagger}$ also remains below
the nominal level across these scenarios, indicating coverage control for the projection-based target as well.

Specifically, as illustrated in Figure~1(a), while the CI lengths of competing methods expand significantly as the number of variables $p$ increases, TwinPoSI maintains consistently shorter confidence intervals. This demonstrates that TwinPoSI provides more precise inference in high-dimensional settings. In Figure~\ref{fig:settings2-3}(b), where the autocorrelation $\rho$ increases, the CI length of TwinPoSI grows more slowly than those of competing methods, demonstrating its robustness to high correlations among variables. In Figure~\ref{fig:settings2-3}(c), as the nominal level $q$ varies, TwinPoSI maintains both empirical FCR and
$\operatorname{FCR}^{\dagger}$ below $q$ and produces shorter CIs than the competing methods across all values considered.

{Appendix}~H considers two additional data-generating schemes: random assignment of 
the nonzero coefficients in \(\bbeta^*\) and the inclusion of interaction terms. 
The corresponding results are shown in Figures~S1 and 
S2, respectively. TwinPoSI exhibits similar empirical advantages under both additional
data-generating schemes, suggesting that its performance is not sensitive to
the particular signal locations or to the inclusion of the interaction terms
considered here.

We further assess robustness to deviations from Gaussian errors by replacing the normal errors in Simulation Setting~\ref{setting:2}(a) with Student's $t$ errors with 3 degrees of freedom. As shown in Figure~\ref{fig:settings2-3}(d), TwinPoSI maintains both empirical FCR and $\operatorname{FCR}^{\dagger}$ below the nominal level while producing shorter confidence intervals and higher efficient power than data splitting and data fission. In contrast, data splitting and data fission exhibit inflated population-based FCR, while their $\operatorname{FCR}^{\dagger}$ values generally remain near or below the nominal level, although data fission slightly exceeds it in some scenarios. Additional results under $t_{10}$, Laplace, and centered asymmetric Gamma errors are reported in Appendix~G.3.

Finally, we examine a weak-signal regime in which support recovery is imperfect,
with $\cS\nsubseteq\hat{\cS}$ occurring with non-negligible probability.
As reported in Appendix~G.4, TwinPoSI continues to control the
population-based FCR near or below the nominal level, consistent with its
theoretical guarantee, which does not require exact support recovery.

\subsection{Numerical Results for Example~\ref{ex:proj-mislead}}

\label{sec:target-mismatch-simu}


We investigate the inferential consequences of Example~\ref{ex:proj-mislead} by comparing six PoSI procedures over 100 independent repetitions of its data-generating model, with $n=1500$, $p=303$, and nominal level $q=0.1$. Implementation details are provided in Appendix~G.5.

We compare six post-selection inference procedures over 100 independent repetitions at the nominal level $q=0.1$. TwinPoSI and BY-dLASSO target population-based coefficients, whereas data splitting, data fission, UPoSI, and EPoSI target projection-based coefficients. For every method, we report both population-based FCR and projection-based $\operatorname{FCR}^{\dagger}$. We also report the proportions of confidence intervals for $\bX_1$ lying entirely below zero, containing zero, or lying entirely above zero, calculated over the repetitions in which $\bX_1$ is selected. These proportions summarize the directional conclusions associated with the different targets rather than provide a common-target power comparison.

As shown in Table~\ref{tab:projection-method-comparison}, data splitting and data fission produce intervals entirely below zero in $52\%$ and $55\%$ of the repetitions in which $\bX_1$ is selected, respectively, consistent with the negative projection-based coefficient induced by omitted correlated covariates. In contrast, BY-dLASSO and TwinPoSI produce intervals entirely above zero in $56\%$ and $94\%$ of selected repetitions, respectively, consistent with $\beta_1^*=1$. UPoSI produces intervals containing zero in all selected repetitions, with an average interval length of $6.398$. EPoSI produces intervals containing zero in $84.7\%$ of selected repetitions and entirely below zero in $14.3\%$, with an average interval length of $2.067$.
\begin{table}
\centering
\caption{Comparison of post-selection inference procedures under Example~\ref{ex:proj-mislead} over 100 repetitions at $q=0.1$. Directional proportions for $\bX_1$ are calculated over repetitions in which it is selected. Bold FCR entries indicate values below the nominal level.}
\label{tab:projection-method-comparison}
\small
\setlength{\tabcolsep}{5pt}
\begin{tabular}{lrrrrrr}
\toprule
Method
& Entirely $<0$
& Contains $0$
& Entirely $>0$
& FCR
& FCR$^\dagger$
& CI length \\
\midrule
TwinPoSI
& 0.060
& 0.000
& 0.940
& \textbf{0.087}
& \textbf{0.086}
& 0.138 \\
BY-dLASSO
& 0.000
& 0.440
& 0.560
& \textbf{0.080}
& 0.115
& 0.321 \\
Data splitting
& 0.520
& 0.480
& 0.000
& 0.150
& \textbf{0.094}
& 0.244 \\
Data fission
& 0.550
& 0.450
& 0.000
& 0.156
& 0.100
& 0.240 \\
UPoSI
& 0.000
& 1.000
& 0.000
& \textbf{0.000}
& \textbf{0.000}
& 6.398 \\
EPoSI
& 0.143
& 0.847
& 0.010
& 0.137
& \textbf{0.086}
& 2.067 \\
\bottomrule
\end{tabular}
\end{table}
TwinPoSI maintains both empirical FCR and $\operatorname{FCR}^{\dagger}$ below the nominal level, at $0.087$ and $0.086$, respectively, while yielding the shortest average interval length of $0.138$. BY-dLASSO has population-based FCR of $0.080$, but its $\operatorname{FCR}^{\dagger}$ is $0.115$. Conversely, data splitting, data fission, and EPoSI maintain $\operatorname{FCR}^{\dagger}$ at or below $0.1$, while their population-based FCR values exceed the nominal level. These results illustrate that procedures can support opposite directional conclusions while providing adequate empirical coverage for their respective targets. In particular, coverage for a projection-based coefficient does not necessarily translate into coverage for the corresponding population-based coefficient.

\subsection{Performance Across Selection Methods}
\label{sec:selection-methods}

We examine whether TwinPoSI's empirical performance extends beyond LASSO by replacing the estimation procedure $\mathcal E$ with SCAD or Elastic Net in both the original fit and the construction and calibration twins. The data-generating settings remain the same as in Simulation Setting~\ref{setting:2}(a). For each selection method, we compare TwinPoSI with data splitting, data fission, and UPoSI over 100 repetitions at $q=0.1$. We omit the EPoSI implementations used in the preceding comparisons because they do not directly accommodate SCAD selection.
\begin{figure}
    \centering
\includegraphics[width=1\linewidth]{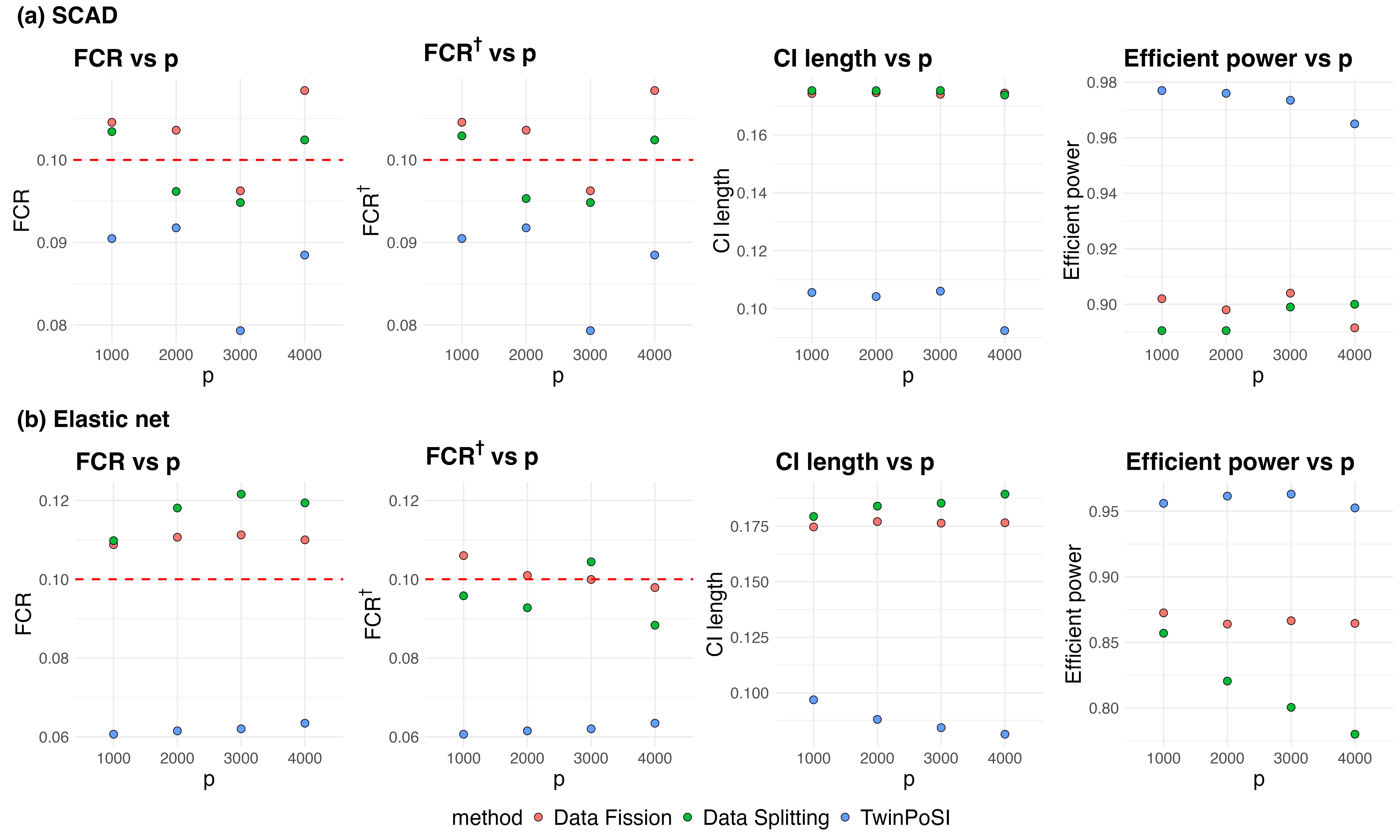}
    \caption{
FCR, FCR$^\dagger$, CI length, and efficient power as the dimension $p$
varies under (a) SCAD selection and (b) Elastic Net selection.
Results are averaged over 100 repetitions, and the red dashed lines
indicate the nominal level $q=0.1$.
}
    \label{fig:selection_methods}
\end{figure}
Figure~\ref{fig:selection_methods} shows that TwinPoSI maintains both empirical FCR and $\operatorname{FCR}^{\dagger}$ below the nominal level under SCAD and Elastic Net selection, while achieving the highest efficient power and shortest average intervals among the methods shown across all dimensions. Under SCAD selection, data splitting and data fission yield similar FCR and $\operatorname{FCR}^{\dagger}$ values near the nominal level, with some values slightly exceeding it. Under Elastic Net selection, their population-based FCR values exceed the nominal level, whereas their $\operatorname{FCR}^{\dagger}$ values generally remain near or below it. These results support TwinPoSI's empirical applicability beyond LASSO, although the theoretical verification remains specific to LASSO.

\section{Real Data Analysis}
\label{sec:real}
We illustrate TwinPoSI using the riboflavin production microarray dataset
for \textit{Bacillus subtilis}, available in the R package \texttt{hdi}
\citep{Dezeure2015hdi}. The dataset contains $n=71$ observations and
$p=4088$ gene-expression variables. The response is the log-transformed
riboflavin production rate, and the covariates are the log-transformed
expression levels of the corresponding probe sets. Following previous
studies \citep{van2014asymptotically,buhlmann2014high}, we fit a linear
regression model to identify genes associated with riboflavin production
and quantify their conditional associations.

We compare TwinPoSI with data fission, data splitting, UPoSI, and EPoSI
at the nominal level $q=0.1$. Procedures involving randomization are
repeated 100 times on the same dataset. We report the average confidence
interval length and the number of significant features, where a selected
feature is declared significant if its interval excludes zero. We also
examine the stability of the inferred association directions across
repeated runs. An interval lying entirely above or below zero indicates
a significant positive or negative association, respectively, with
respect to the coefficient targeted by the method.

Table~\ref{tab:ribo_ci_sig_compare} summarizes the results. TwinPoSI
produces the shortest average finite interval length, 0.17, and the
largest average number of significant features, 8.62. Data fission and
data splitting yield average lengths of 0.48 and 0.63, with 3.95 and
2.94 significant features, respectively. EPoSI has an average
interval length of 27.47 and occasionally returns excessively lengthy intervals,
while identifying 1.84 significant features on average. UPoSI yields
an average finite interval length of 38.79 and no significant features.
The features identified by TwinPoSI include the
\texttt{YXLD\_at} and \texttt{YOAB\_at} probe sets,
corresponding to the \textit{YXLD} and \textit{YOAB} genes,
respectively, which were also reported in previous analyses
\citep{buhlmann2014high}.

Figure~S6 displays the intervals for
the \texttt{YXLD\_at} probe set across repeated runs. TwinPoSI
returns intervals lying entirely below zero in all 100 runs,
showing that its negative association finding is stable with
respect to algorithmic randomization on this dataset. This negative association
agrees with the analysis of the same dataset by
\cite{fei2019drawing}. Data fission and data splitting select \texttt{YXLD\_at} less
frequently and generally produce wider intervals. Data splitting
and EPoSI also return intervals lying entirely above zero in some
runs, whereas UPoSI produces intervals that include zero, which shows that TwinPoSI gives a more stable directional
conclusion for this probe set across repeated runs.

The identified genes provide several biological leads.
The negative association for \texttt{ARGF\_at}, which encodes
ornithine carbamoyltransferase involved in arginine biosynthesis,
suggests a possible link between amino acid metabolism and
riboflavin production. The positive associations for
\texttt{XHLB\_at} and \texttt{SPOVAA\_at} involve genes associated
with PBSX-mediated cell lysis and spore formation, respectively (Figure~S6).
These findings raise the possibility that variation in production
rate is associated with broader differences in cellular physiology.
They motivate further investigation of these processes, but do not
establish that altering the identified genes would change
riboflavin production.
\begin{table*}[t]
\centering
\small
\caption{
Confidence interval lengths and numbers of significant features
for the riboflavin dataset. Entries are means with standard errors
in parentheses over 100 runs. Bold values indicate the smallest average
finite interval length and the largest average number of significant
features.
}
\label{tab:ribo_ci_sig_compare}
\begin{tabular}{lrrrrr}
\toprule
& TwinPoSI & Data Fission & Data Splitting & UPoSI & EPoSI \\
\midrule
CI Length & \textbf{0.17 (0.00)} & 0.48 (0.01) & 0.63 (0.02) & 38.79 (0.40) & 27.47 (6.34) \\
\# Significant Features & \textbf{8.62 (0.22)} & 3.95 (0.20) & 2.94 (0.18) & 0.00 (0.00) & 1.84 (0.44) \\
\bottomrule
\end{tabular}
\end{table*}

\section{Discussion}
We introduced TwinPoSI, a statistical-twin-based method for post-selection inference on population-based regression coefficients. By calibrating confidence interval expansion using synthetic datasets generated from a fitted model, TwinPoSI controls the false coverage rate up to a Monte Carlo error and an asymptotically negligible approximation error. Under a Gaussian linear model with an orthogonal design, TwinPoSI is the only method considered that achieves first-order oracle efficiency in average interval length. Simulations show that TwinPoSI achieves reliable FCR control while producing shorter confidence intervals and higher efficient power than competing methods.

The distinction between population-based and projection-based targets is important for scientific interpretation. Projection coefficients depend on the selected sub-model and may absorb the contributions of omitted correlated variables, whereas population-based coefficients refer to the specified full model. However, population-based coefficients remain model-dependent conditional associations and should not be interpreted as causal effects without additional assumptions. Because the competing procedures considered here do not all target the same coefficient definition, comparisons of confidence interval length and numbers of significant findings in the real-data analyses should be viewed as comparisons of practical inferential outputs rather than strict common-target efficiency comparisons.

Several limitations warrant further investigation. First, statistical twins are generated from a fitted parametric model, and their ability to calibrate post-selection uncertainty depends on how well this model approximates the original data-generating distribution. Although our simulations suggest robustness to several departures from Gaussianity, a systematic theoretical analysis under model misspecification remains an important direction. Second, the FCR guarantee requires that statistical twins
do not systematically underestimate the original noncoverage
risk. Although we provide sufficient conditions for this
assumption under LASSO, assessing whether it holds in a given
application remains challenging. Developing practical
diagnostics for this approximation would help identify
settings in which TwinPoSI can be expected to perform reliably. Third, TwinPoSI widens both ends of each initial interval by the same
amount. Allowing different amounts of widening at the lower and upper
endpoints may yield shorter intervals when estimation errors are
asymmetric. Such an extension would require a calibration rule for
the two widening parameters and a corresponding FCR guarantee.

Our work also opens several broader directions for future research. Adapting TwinPoSI to generalized linear models, Gaussian graphical models, mixed-effects models, and other complex modeling frameworks would substantially broaden its applicability. Another direction is to develop feature-specific or scale-adaptive interval adjustments. Rather than applying a common additive correction factor $\hat{T}_q$ to all selected coefficients, feature-dependent corrections could account for differences in standard errors, signal strengths, and selection frequencies, potentially improving interval efficiency while preserving FCR control. Extending TwinPoSI to other post-selection problems, such as change-point detection and causal discovery, would further expand its scope beyond regression-based feature selection \citep{hyun2021post,chang2026post}. Finally, applying TwinPoSI to additional scientific problems, including expression quantitative trait locus (eQTL) analysis, would provide further evidence of its practical utility \citep{westra2013systematic}.
\bibliographystyle{chicago}
\bibliography{TwinPoSI.bib}



\end{document}